\documentclass[11pt,letterpaper]{article}
\usepackage[T1]{fontenc}
\usepackage[utf8]{inputenc}
\usepackage{lmodern}
\usepackage[margin=1in]{geometry}
\usepackage{amsmath,amssymb}
\usepackage{graphicx}
\usepackage{booktabs}
\usepackage{microtype}
\usepackage{placeins}
\usepackage{xurl}
\usepackage[hidelinks]{hyperref}
\hypersetup{pdftitle={Logistic vegetation trajectories and their implications for land surface phenology},pdfauthor={Hongxiao Jin}}
\title{Logistic vegetation trajectories and their implications for land surface phenology}
\author{Hongxiao Jin\\[0.4em]\small Department of Earth and Environment Science, Lund University, Sweden}
\date{}
\begin{document}
\maketitle
\begin{abstract}

Vegetation trajectories describe canopy development and provide the
basis for retrieving phenological metrics, including the start and end
of the growing season. Here, combining exponential development with a
Michaelis--Menten-type saturating response, we derive a generalized
logistic function (GLF) to represent canopy growth trajectories under
explicit developmental and response assumptions. Using this function as
a common mathematical basis, we establish relationships between
trajectory shape, phenological metrics and date uncertainty. We derive
shape-dependent relative thresholds for derivative-defined transition
points and recover the 9.18\% onset threshold as a special case for an
isolated ordinary logistic growth curve under the small-slope
approximation. For a symmetric double-logistic seasonal trajectory, we
derive the realized amplitude, full width at half maximum and a critical
dimensionless growth curve separation at which interior third-derivative
extrema merge. We further show that transition points defined by exact
curvature-change extrema depend on the metric assigned to the time and
response axes. Local noise propagation and a reproducible experiment
comprising 4,800 nonlinear fits connect trajectory geometry with date
uncertainty, demonstrating that additional shape flexibility can reduce
bias while increasing uncertainty in the seasonal tails. Together, the
GLF construction and its analytical extensions explain how trajectory
shape determines the location and existence of phenological transition
points and the uncertainty in their estimated dates. This framework
supports phenology retrieval that preserves the underlying trajectory
and interprets each metric through its mathematical definition and
uncertainty.

\end{abstract}


\section{Introduction}\label{introduction}

Satellite time series describe the course of vegetation development,
while land surface phenology products commonly summarize that course
using transition dates. The relationship between the continuous
trajectory and its summaries is central to interpreting vegetation
responses across years, locations and observation systems. Threshold
crossings, maximum slope and extrema of curvature change describe
distinct stages of a fitted curve. Understanding their relationships
helps connect convenient date summaries with the dynamics from which
they are extracted.

Logistic functions have a long history in satellite phenology, including
the formulation of Zhang et al. (2003) and later hybrid models (Zhang,
2015; Zhang et al., 2018). Shang et al. (2017) established the
correspondence between derivative landmarks and relative thresholds for
an ordinary logistic growth curve. Using a small-slope approximation to
curvature, they derived the 9.18\% onset threshold and reported the
21.13\% and 50\% landmarks. They also demonstrated how growth rate
affects differences between threshold dates and illustrated extraction
from mixed vegetation and a water-stressed trajectory. These results
provide an analytical foundation for extending the correspondence to a
wider range of trajectory shapes.

Asymmetry and uncertainty have also been examined empirically.
Klosterman et al. (2014) fitted generalized sigmoid curves, observed
shifts in curvature-change landmarks relative to fixed amplitude
percentages, and estimated date uncertainty by sampling from fitted
parameter covariance. Verma et al. (2016) compared logistic, Richards
and Gompertz-type growth functions for remotely sensed forest phenology.
The remaining opportunity is to quantify how curve shape determines the
threshold associated with a landmark, how seasonal overlap changes the
existence of landmarks, and how these geometric properties affect date
estimation.

Here we develop that analysis from an explicit construction of logistic
growth to the geometry and estimation of phenological landmarks. We
derive threshold relationships for asymmetric growth curves, distinguish
exact curvature from its third-derivative approximation, and analyse the
interaction of growth and decline in a composite seasonal curve. A
synthetic fitting experiment then examines the consequences of
estimating asymmetry for date bias and uncertainty. The analysis
connects three questions: where a landmark occurs, whether it exists,
and how precisely it can be estimated.

\section{Construction of logistic vegetation
trajectories}\label{construction-of-logistic-vegetation-trajectories}

\subsection{Exponential development and a saturating
response}\label{exponential-development-and-a-saturating-response}

Consider an intermediate quantity \(S(t) > 0\) that increases
exponentially over a restricted developmental interval. With a constant
rate \(k > 0\) and initial value \(S_{0} > 0\),

\begin{equation}
\label{eq:1}
\frac{dS}{dt} = kS,\quad\quad S(t) = S_{0}\exp(kt).
\end{equation}

If \(S\) represents a cell population with doubling time \(T_{d}\), then
\(k = \log(2)/T_{d}\). Thus \(2^{t}\) expresses exponential
multiplication when time is measured in doubling-time units. Over a
developmental interval with approximately constant division rate, this
provides a simple representation of proliferative potential. Cell
expansion, turnover and changing resource supply can subsequently be
incorporated when linking that potential to canopy development.

Suppose the observable response \(V\) is a rectangular-hyperbolic
function of this quantity, with fixed maximum \(V_{\max}\) and
half-saturation constant \(K > 0\),

\begin{equation}
\label{eq:2}
\begin{aligned}
& V(t) = V_{\max}\frac{S(t)}{K + S(t)} = \frac{V_{\max}}{1 + \exp\{ - k(t - b)\}},\\
& b = \frac{\log\left( K/S_{0} \right)}{k}.
\end{aligned}
\end{equation}

Substitution of (1) into the Michaelis--Menten-type response gives the
ordinary logistic curve in (2), with dimensionless ratio \(K/S_{0}\).
The construction links exponential development and saturation through an
explicit mathematical relationship. Here the rectangular-hyperbolic
response is used as a representation of saturation, motivated by
Michaelis and Menten (1913); its canopy interpretation depends on the
response variable. Related work has combined logistic plant-growth
models with photosynthetic saturation and cellular processes (Kawano et
al., 2020).

The intermediate quantity can represent developmental potential or a
structural driver of the measured response. This interpretation
separates cell proliferation from the optical or physiological quantity
being modelled. In canopy applications, the response may depend on green
leaf area, light interception or photosynthetic capacity. Leaf-level
biochemical limitations and canopy light distribution provide further
physiological structure for such applications (Farquhar et al., 1980).

The choice of saturation function determines the resulting trajectory
family. For example, applying \(V/V_{\max} = 1 - \exp( - \beta S)\) to
(1) gives \(1 - \exp\{ - \beta S_{0}\exp(kt)\}\), a complementary
Gompertz form. Exponential development therefore provides a common
starting point, while the response to \(S\) distinguishes alternative
sigmoid models.

\subsection{Generalized logistic form}\label{generalized-logistic-form}

Write an isolated increasing response as

\begin{equation}
\label{eq:3}
\begin{aligned}
& x(t) = c + AG(t),\\
& G(t) = \{ 1 + \exp\left\lbrack - k(t - b) \right\rbrack\}^{- \alpha},\quad\quad A,k,\alpha > 0.
\end{aligned}
\end{equation}

Here \(c\) is the baseline, \(A\) the difference between asymptotes,
\(b\) a location parameter, and \(\alpha\) a shape parameter controlling
asymmetry. Equation (3) belongs to the Richards family (Richards, 1959),
with the ordinary logistic curve recovered at \(\alpha = 1\). The
extension introduces \(\alpha\) by raising the normalized saturating
response in (2) to the power \(\alpha\). This additional response
assumption allows different rates of approach to the lower and upper
asymptotes.

Differentiating (3) gives an equivalent autonomous equation,

\begin{equation}
\label{eq:4}
\frac{dG}{dt} = \alpha kG\left( 1 - G^{1/\alpha} \right).
\end{equation}

Equation (4) expresses the generalized trajectory as an autonomous
growth equation. Its early relative growth rate tends to \(\alpha k\),
while approach to the upper asymptote is controlled by \(k\). These two
rates give the shape parameter a direct mathematical interpretation and
provide a basis for comparing early development with later saturation.

\subsection{Decline and the observation
process}\label{decline-and-the-observation-process}

A declining growth curve can be represented by \(c + A\{ 1 - G(t)\}\) or
by the time-reversed curve \(c + AG( - t)\), with appropriate location
parameters. These constructions describe different asymmetric decline
patterns when \(\alpha \neq 1\). Their selection allows senescence and
leaf loss to have dynamics distinct from canopy expansion, consistent
with the role of leaf production and turnover in canopy development
(Hikosaka, 2005).

Let the underlying canopy variable be \(L(t)\) and the observed index be
\(x(t) = h\{ L(t),z(t)\}\), where \(z\) represents additional
environmental or structural influences. The relationship between a
logistic \(L(t)\) and the observed \(x(t)\) depends on the form and
temporal behaviour of \(h\). Accordingly, the following analysis is
expressed in terms of the fitted response \(x\), so that the same
mathematical framework can be evaluated separately for foliage, optical
greenness and carbon assimilation.

\section{Transition functionals and their
relationships}\label{transition-functionals-and-their-relationships}

\subsection{Relative thresholds and maximum
slope}\label{relative-thresholds-and-maximum-slope}

For a fixed relative threshold \(p \in (0,1)\), define \(t_{p}\) by
\(x\left( t_{p} \right) = c + pA\). Inverting (3) gives

\begin{equation}
\label{eq:5}
t_{p} = b - \frac{1}{k}\log\left( p^{- 1/\alpha} - 1 \right).
\end{equation}

The threshold date is an explicit function of the curve parameters.
Conversely, each point on a strictly increasing growth curve has an
equivalent relative threshold. This correspondence, established for
ordinary logistic landmarks by Shang et al. (2017), provides the basis
for quantifying how landmark thresholds vary with asymmetry.

Put \(q = \{ 1 + \exp\left\lbrack - k(t - b) \right\rbrack\}^{- 1}\), so
that \(G = q^{\alpha}\). The slope is \(A\alpha kq^{\alpha}(1 - q)\) and
is largest at \(q = \alpha/(\alpha + 1)\). Consequently,

\begin{equation}
\label{eq:6}
\begin{aligned}
& t_{slope} = b + \frac{\log\alpha}{k},\\
& p_{slope} = \left( \frac{\alpha}{\alpha + 1} \right)^{\alpha}.
\end{aligned}
\end{equation}

For the ordinary logistic curve, the location parameter \(b\),
half-amplitude date and maximum-slope date coincide. With
\(\alpha = 0.4\), maximum slope occurs at 60.59\% of amplitude. The
location parameter \(b\), the half-amplitude date and the maximum-slope
date therefore provide distinct descriptions of an asymmetric
trajectory.

\subsection{Inflection and curvature
change}\label{inflection-and-curvature-change}

An ordinary inflection of \(x(t)\) satisfies \(x''(t) = 0\) and marks
maximum slope. The earlier curvature-change landmark used for onset
represents a different derivative feature. For an ordinary logistic
growth curve, extrema of \(x''\) occur at
\(\left( 3 \pm \sqrt{3} \right)/6\), approximately 21.13\% and 78.87\%;
the early value corresponds to the curvature landmark reported by Shang
et al. (2017) under their small-slope approximation.

In selected numerical coordinates, signed geometric curvature and its
time derivative are

\begin{equation}
\label{eq:7}
\begin{aligned}
& \kappa(t) = \frac{x''}{\{ 1 + (x')^{2}\}^{3/2}},\\
& \dot{\kappa}(t) = \frac{x'''}{\{ 1 + (x')^{2}\}^{3/2}} - \frac{3x'(x'')^{2}}{\{ 1 + (x')^{2}\}^{5/2}}.
\end{aligned}
\end{equation}

When the numerical slopes are sufficiently small, the denominators
approach one and \(\dot{\kappa}\) is approximated by \(x'''\). Extrema
of this approximation satisfy \(x^{(4)} = 0\). For an isolated ordinary
logistic curve,

\begin{equation}
\label{eq:8}
x^{(4)} = Ak^{4}q(1 - q)\left( 1 - 14q + 36q^{2} - 24q^{3} \right).
\end{equation}

The three interior roots are

\begin{equation}
\label{eq:9}
\begin{aligned}
& q_{-} = \frac{3 - \sqrt{6}}{6},\\
& q_{0} = \frac{1}{2},\quad\quad q_{+} = \frac{3 + \sqrt{6}}{6}.
\end{aligned}
\end{equation}

The early maximum of \(x'''\) occurs at 9.175\% of amplitude, the
central minimum at 50\%, and the late maximum at 90.825\%. Equations
(8)--(9) reproduce the ordinary-logistic correspondence of Shang et al.
(2017) and establish the reference case for the generalized analysis
below. In derivative terminology, the onset landmark is a local maximum
of the third derivative under the small-slope approximation.

For generalized logistic curves, define polynomials \(P_{n}(q)\)
recursively by

\begin{equation}
\label{eq:10}
\begin{aligned}
& P_{0}(q) = 1,\\
& P_{n + 1}(q) = \alpha(1 - q)P_{n}(q) + q(1 - q)P_{n}'(q).
\end{aligned}
\end{equation}

Then \(G^{(n)} = k^{n}q^{\alpha}P_{n}(q)\). The real roots of \(P_{4}\)
in \((0,1)\) locate stationary points of \(G'''\), with equivalent
thresholds \(p = q^{\alpha}\). Classification of these points identifies
the early maximum, central minimum and late maximum. Table 1 and Figure
1 quantify the shape dependence anticipated by the empirical
observations of Klosterman et al. (2014): the early threshold ranges
from 28.55\% at shape 0.25 to 2.91\% at shape 4. The fixed 9.18\%
correspondence is recovered at shape 1.

\begin{table}[htbp]
\centering
\caption{Equivalent thresholds for isolated generalized logistic growth
curves. The first three columns of results describe extrema of the third
derivative, used here as a small-slope approximation to curvature-change
extrema. Maximum slope is defined separately by (6).}
\label{tab:thresholds}
\small
\begin{tabular}{ccccc}
\toprule
Shape $\alpha$ & Early maximum & Central minimum & Late maximum & Maximum slope \\
\midrule
0.25 & 0.28549 & 0.74367 & 0.95773 & 0.66874 \\
0.40 & 0.20182 & 0.66106 & 0.94220 & 0.60586 \\
0.50 & 0.16842 & 0.62033 & 0.93410 & 0.57735 \\
1.00 & 0.09175 & 0.50000 & 0.90825 & 0.50000 \\
2.00 & 0.04959 & 0.40483 & 0.88558 & 0.44444 \\
4.00 & 0.02908 & 0.34130 & 0.86927 & 0.40960 \\
\bottomrule
\end{tabular}
\end{table}

\begin{figure}[!htbp]
\centering
\includegraphics[width=\textwidth]{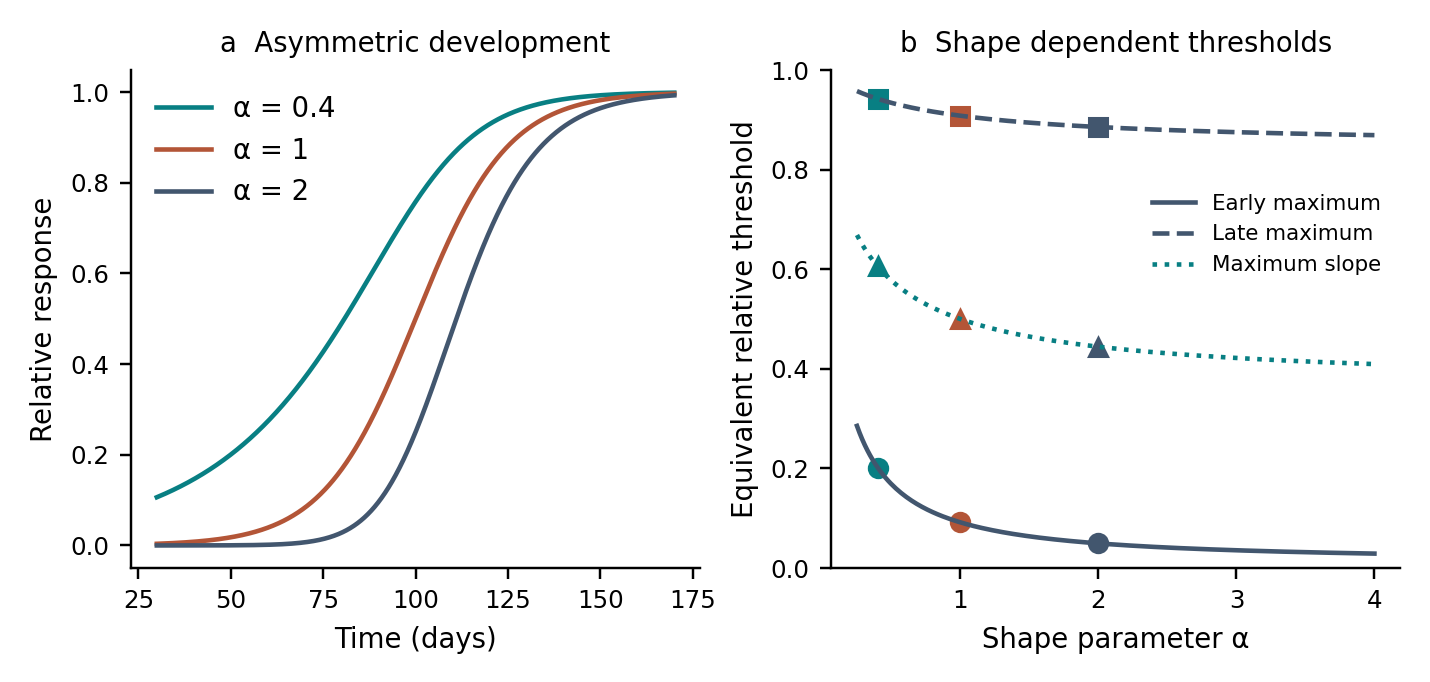}
\caption{Generalized logistic shape and derivative-defined transitions.
Panel a holds location and rate fixed while varying asymmetry. Panel b
shows equivalent relative thresholds for extrema of the third derivative
and for maximum slope.}
\label{fig:1}
\end{figure}

\subsection{Dependence of exact curvature on coordinate
scaling}\label{dependence-of-exact-curvature-on-coordinate-scaling}

Geometric curvature assigns a metric to both axes. Let \(\tau = t/T\)
and \(y = (x - c)/H\), where \(T\) and \(H\) are selected time and
response scales. For an ordinary logistic curve, the numerical slope of
\(y\) against \(\tau\) contains the factor \(\lambda = AkT/H\). Exact
curvature-change extrema depend on this factor. In contrast, the
locations of third-derivative extrema are invariant under positive
affine rescaling of time and response.

Numerical evaluation of (7) gives early-transition relative levels of
0.09175, 0.08747 and 0.02875 for \(\lambda = 0.01\), 1 and 10,
respectively (Figure 2a). Rescaling the axes changes numerical
\(\lambda\) unless their metric is held fixed. The third-derivative
approximation thus provides a scale-invariant reference, while exact
curvature requires the time and VI scales to be specified. This
identifies the mathematical condition under which the fixed-threshold
approximation of Shang et al. (2017) remains close to exact curvature
change.

\section{Seasonal overlap and transition
availability}\label{seasonal-overlap-and-transition-availability}

A seasonal curve can be constructed by subtracting two generalized
sigmoid growth curves. To isolate overlap, consider the simpler
symmetric ordinary-logistic construction, without an additional
declining baseline,

\begin{equation}
\label{eq:11}
\begin{aligned}
& x(t) = c + A\{ L\left\lbrack k\left( t - b_{g} \right) \right\rbrack - L\left\lbrack k\left( t - b_{s} \right) \right\rbrack\},\\
& L(u) = \frac{1}{1 + e^{- u}},\\
& D = b_{s} - b_{g} > 0.
\end{aligned}
\end{equation}

The maximum occurs at \(m = \left( b_{g} + b_{s} \right)/2\). Writing
\(u = k(t - m)\) and \(r = kD\), the normalized seasonal response is

\begin{equation}
\label{eq:12}
\begin{aligned}
& \frac{x(t) - c}{A} = \frac{\sinh(r/2)}{\cosh(u) + \cosh(r/2)},\\
& \frac{x(m) - c}{A} = \tanh(r/4).
\end{aligned}
\end{equation}

Thus \(A\) is a latent growth curve amplitude, not the realized seasonal
amplitude when the growth curves overlap. At \(r = 2\), 4 and 8, the
realized fractions are 0.4621, 0.7616 and 0.9640. The full width at half
the realized maximum is

\begin{equation}
\label{eq:13}
W_{1/2} = \frac{2}{k}\operatorname{arcosh}\{ 2 + \cosh(r/2)\}.
\end{equation}

The separation \(D\) is therefore not exactly the full width at half
maximum. For \(r = 2\), the dimensionless width \(kW_{1/2}\) is 3.8752,
appreciably larger than \(kD = 2\).

Overlap also changes derivative geometry. For the composite seasonal
curve, \(x'''\) is antisymmetric about \(m\). Its central stationary
condition is \(x^{(4)}(m) = 2Ak^{4}L^{(4)}(r/2) = 0\), where the last derivative
is with respect to the dimensionless argument. The nonzero critical
separation is

\begin{equation}
\label{eq:14}
r_{c} = 2\log\left( 5 + 2\sqrt{6} \right) \simeq 4.58486.
\end{equation}

The analytic derivatives show how the transition sequence changes with
separation. At \(r = 8\), the rising half contains two positive local
maxima of \(x'''\), representing distinct early and interior landmarks;
at \(r = 2\) or 4, only the early positive maximum remains (Figure 2b).
The interior maximum and its falling-side counterpart merge at the
seasonal centre at the critical separation. Equation (14) therefore
identifies a change in landmark existence for the composite curve in
(11). When selecting extrema numerically, an interior-stationarity check
distinguishes a resolved landmark from an interval boundary.

For every \(p \in (0,1)\), a threshold referenced to the realized
seasonal peak remains uniquely defined on each strictly monotone half of
(11), including separations for which the interior derivative landmarks
have merged. Thresholds referenced to the latent amplitude \(A\) have a
different domain of availability because the composite trajectory may
remain below the target level. Amplitude reference and growth curve
interaction therefore jointly determine which date summaries are
available.

The overlap result applies to differentiation of the composite seasonal
curve. The VIIRS guide and ATBD describe fitting increasing and
decreasing phases separately (VIIRS GLSP Science Team, 2016, 2020).
Phase-wise fitting and composite fitting offer different representations
of growth curve interaction. Their practical comparison can use the
separation criterion as a diagnostic while retaining each algorithm's
segmentation and transition-selection rules.

\begin{figure}[!htbp]
\centering
\includegraphics[width=\textwidth]{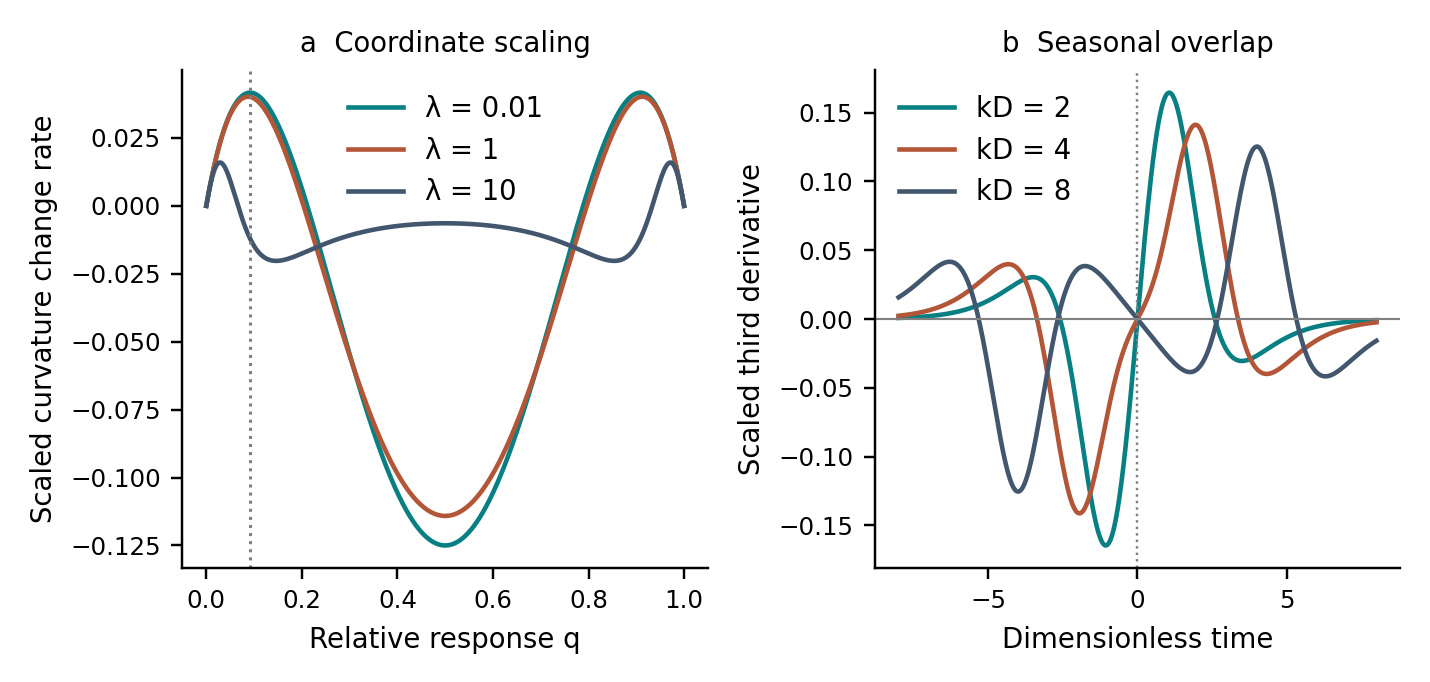}
\caption{Coordinate scaling and seasonal overlap. Panel a evaluates
exact curvature-change rate for an isolated ordinary logistic growth
curve at different numerical slope scales \(\lambda\); curves are divided by their
common-form amplitude factor for display. Panel b shows the analytic
third derivative of the composite seasonal curve for three dimensionless
separations.}
\label{fig:2}
\end{figure}

\section{Estimation and uncertainty}\label{estimation-and-uncertainty}

\subsection{Local noise equivalent date
uncertainty}\label{local-noise-equivalent-date-uncertainty}

If the target level \(x_{*}\) is fixed and a crossing is locally
monotone, a small response perturbation \(\delta x\) gives

\begin{equation}
\label{eq:15}
\delta t \simeq \frac{\delta x}{x'\left( t_{*} \right)},\quad\quad\sigma_{t} \simeq \frac{\sigma_{x}}{\left| x'\left( t_{*} \right) \right|}.
\end{equation}

Equation (15) converts response uncertainty into date uncertainty at a
fixed, locally monotone crossing with nonzero slope. It isolates the
contribution of local curve geometry. For (3), the slope at relative
threshold \(p\) is

\begin{equation}
\label{eq:16}
\left| x'\left( t_{p} \right) \right| = A\alpha kp\left( 1 - p^{1/\alpha} \right).
\end{equation}

With equal local VI uncertainty, noise-equivalent date uncertainty is
smallest at the maximum-slope level in (6). For an ordinary logistic
curve, uncertainty at 10\% amplitude is \(0.25/(0.1 \times 0.9) = 2.78\)
times that at 50\%. This quantifies local sensitivity; uncertainty in a
fitted transition additionally reflects parameter estimation. Where VI
uncertainty varies through the season, the relevant ratio is
\(|x'|/\sigma_{x}\). Biological accuracy can then be evaluated by
relating the selected crossing to independently observed canopy
development.

\subsection{Uncertainty in a fitted transition
functional}\label{uncertainty-in-a-fitted-transition-functional}

Let \(Y_{i} = x\left( t_{i};\theta \right) + \varepsilon_{i}\), with
parameter estimate \(\widehat{\theta}\) and covariance
\({\widehat{\Sigma}}_{\theta}\). For a smooth date functional
\(T(\theta)\), the delta approximation is

\begin{equation}
\label{eq:17}
\widehat{\operatorname{Var}}\{ T\left( \widehat{\theta} \right)\} \simeq \nabla T\left( \widehat{\theta} \right)^{\mathsf{T}}{\widehat{\Sigma}}_{\theta}\nabla T\left( \widehat{\theta} \right).
\end{equation}

For a crossing defined implicitly by \(x(T;\theta) = C(\theta)\),

\begin{equation}
\label{eq:18}
\nabla_{\theta}T = \frac{\nabla_{\theta}C - \nabla_{\theta}x(T;\theta)}{x'(T;\theta)}.
\end{equation}

The derivative of \(x\) in the numerator holds time fixed. A fixed
absolute threshold has \(\nabla C = 0\); a relative threshold estimated
from the same curve has \(C = c + pA\). The covariance among fitted
parameters therefore matters even when the relative-date formula does
not explicitly contain \(c\) or \(A\). Sparse observations, incomplete
coverage and shape uncertainty can increase uncertainty beyond (15).

For a derivative-defined transition, the same implicit approach uses its
defining derivative equation in place of the threshold crossing. As
landmarks approach a merger, their defining roots become nearly
degenerate and sensitivity increases. Uncertainty can then be described
jointly by the distribution of resolved dates and the probability that a
qualifying landmark exists. This suggests a useful extension of
parameter-based uncertainty estimation, including the
covariance-sampling approach used by Klosterman et al. (2014).

\section{Synthetic fitting
experiment}\label{synthetic-fitting-experiment}

\subsection{Design}\label{design}

We generated isolated increasing trajectories using (3), with
\(c = 0.1\), \(A = 0.7\), \(k = 0.08\,\mathrm{day}^{- 1}\), \(b = 100\) days
and \(\alpha \in \{ 0.4,1,2\}\). Observations were taken from day 0 at
intervals of 4 or 12 days through the last sampling date not exceeding
day 200. Independent Gaussian errors had standard deviation 0.01 or 0.03
VI units. The resulting 12 scenarios each contained 200 replicates,
using NumPy random seed 20261017.

Every replicate was fitted by bounded nonlinear least squares using an
ordinary logistic model and a generalized logistic model with estimated
shape. Both estimated baseline, amplitude, rate and location. Rate and
shape were optimized on logarithmic scales. Bounds were
\(c \in \lbrack - 0.3,0.5\rbrack\), \(A \in \lbrack 0.05,1.5\rbrack\),
\(k \in \lbrack 0.005,0.5\rbrack\), \(b \in \lbrack 0,200\rbrack\) and,
for the generalized model, \(\alpha \in \lbrack 0.1,10\rbrack\). Initial
values were 0.1, 0.7, 0.08 and 100, with shape initialized at 1. Common
starting values provide a controlled comparison of model shape, sampling
interval and noise.

Dates were calculated at \(p = 0.1\), 0.5 and 0.9, with reference values
defined by the generating curve through (5). Standard errors used
residual variance and the pseudoinverse of the least-squares information
matrix in fitting coordinates, followed by numerical differentiation of
(5). Nominal 95\% Wald intervals used
\(\widehat{t} \pm 1.96\widehat{\mathrm{SE}}\). All fits were retained, including
those close to a bound. Bias, root mean squared error (RMSE), interval
coverage and bound attainment together describe estimation performance.

\subsection{Results}\label{results}

All 4,800 fits returned optimizer success. Of these, 135 reached at
least one parameter bound to within \(10^{- 5}\) in optimization
coordinates, indicating scenarios in which the available observations
gave limited support for the fitted parameter combination.

For 12-day sampling and noise standard deviation 0.03, an ordinary
logistic fit to a true shape of 0.4 gave a 10\% date bias of 14.93 days
and RMSE of 16.11 days. Allowing shape to vary reduced the bias to -4.84
days but increased RMSE to 22.22 days; 21 of 200 generalized fits
reached a bound. The corresponding coverage increased from 34.5\% to
84.0\%, still below nominal. The greater flexibility reduced systematic
error while amplifying uncertainty in the early tail.

When the true curve was ordinary logistic under the same sampling and
noise, the ordinary fit gave RMSE values of 3.13, 1.59 and 3.32 days at
the 10\%, 50\% and 90\% dates. Estimating an additional shape parameter
increased these to 6.41, 1.62 and 4.96 days. Conversely, for true shape
2, the 90\% date bias decreased from -1.92 to -0.07 days with estimated
shape, while RMSE increased from 3.71 to 4.50 days. These examples show
why flexibility, bias and precision should be evaluated separately
(Figure 3).

With 200 replicates, coverage near 95\% has a Monte Carlo standard error
of approximately 1.5 percentage points. Full scenario summaries,
individual replicate results, curve calculations and the executable
script accompany the manuscript. These controlled comparisons resolve
how estimating asymmetry affects threshold dates under different
sampling intervals and VI noise levels.

\begin{figure}[!htbp]
\centering
\includegraphics[width=\textwidth]{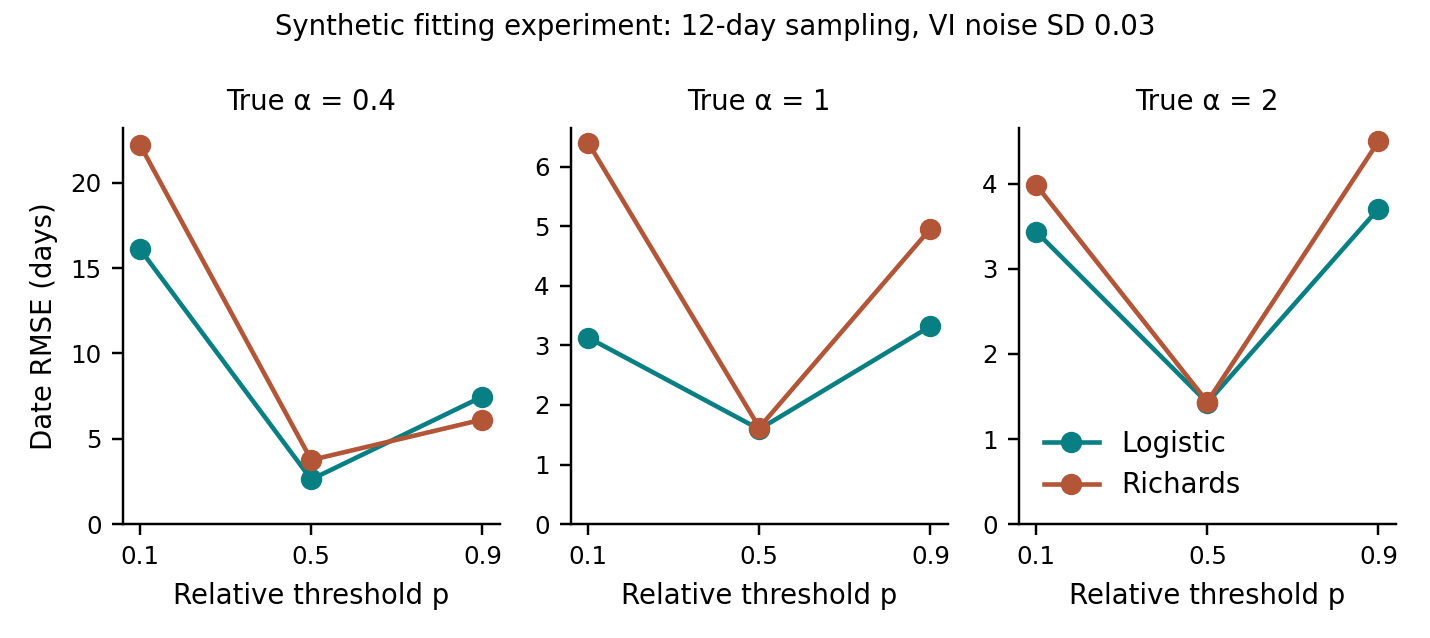}
\caption{Date estimation in the synthetic fitting experiment. The
plotted scenarios use 12-day sampling, VI noise standard deviation 0.03
and 200 replicates per true shape. Both models estimate baseline,
amplitude, rate and location; the Richards model also estimates shape.
RMSE is relative to the generating curve threshold date.}
\label{fig:3}
\end{figure}

\section{Discussion}\label{discussion}

\subsection{Extension of the threshold
correspondence}\label{extension-of-the-threshold-correspondence}

Shang et al. (2017) derived the ordinary-logistic onset threshold of
9.18\%, reported the 21.13\% curvature and 50\% maximum-growth-rate
landmarks, and related differences between threshold dates to growth
rate. Their global NDVI analysis and examples of mixed vegetation and
summer water stress demonstrated applications beyond the idealized
curve. These findings are the starting point of the present work.

The extension developed here concerns the dependence and availability of
landmarks. Equation (10) and Table 1 quantify their threshold dependence
on the generalized-logistic shape parameter. Equations (12)--(14)
connect growth--decline overlap to realized amplitude, half-maximum
width and the merger of interior third-derivative landmarks. Equation
(7) separates exact curvature change from the small-slope approximation
explicitly used by Shang et al. (2017). The estimation analysis then
distinguishes local response sensitivity from fitted-parameter
uncertainty and illustrates the bias--uncertainty trade-off under
controlled sampling.

Asymmetric sigmoid fitting, landmark shifts and parameter-based
uncertainty were already investigated by Klosterman et al. (2014). The
advance developed here is their analytical connection to shape-dependent
thresholds and to the existence of landmarks under seasonal overlap. The
derivations supply quantitative relationships that can be examined
across vegetation indices and retrieval models.

\subsection{Growth assumptions and their
interpretation}\label{growth-assumptions-and-their-interpretation}

The growth construction connects a simple developmental hypothesis to a
tractable family of vegetation trajectories. Exponential proliferation
combined with a rectangular-hyperbolic response yields the ordinary
logistic curve, and a power response extends it to asymmetric
development. The value of this formulation is the explicit link between
assumptions, shape parameters and observable curve properties. Its
canopy interpretation can be developed through measurements that relate
cell proliferation, leaf expansion and the chosen response variable.

This construction complements existing logistic plant-growth and
photosynthesis models (Kawano et al., 2020) and the use of generalized
sigmoid functions in phenology (Klosterman et al., 2014; Verma et al.,
2016). The mathematical growth family provides the foundation; the
present analysis develops its implications for landmark thresholds,
seasonal interaction and date estimation. A physiologically
interpretable trajectory can be extended through separate descriptions
of canopy turnover, resource limitation and the observation process.

\subsection{Choosing and reporting
dates}\label{choosing-and-reporting-dates}

Threshold and derivative summaries offer complementary ways to describe
development. Shang et al. (2017) connected them through fixed
percentages for an ordinary logistic growth curve and demonstrated the
practical value of this correspondence. The generalized analysis
replaces these fixed percentages with shape-dependent relationships. It
also shows that seasonal interaction can change the number of available
landmarks, while exact curvature introduces dependence on the coordinate
metric. Together, these results distinguish a change in the date of a
landmark from a change in the geometry that defines it.

A useful reporting framework therefore includes the response variable,
curve family, shape parameters, amplitude reference, derivative
definition and uncertainty of the selected date. For derivative
landmarks, the status of the qualifying extremum adds information about
whether the developmental stage is geometrically resolved. These
descriptions can support comparisons across products by identifying
compatible quantities and date definitions (Moon et al., 2019).

\subsection{Relation to trajectory based
phenology}\label{relation-to-trajectory-based-phenology}

The trajectory provides the common structure connecting these summaries.
Magnitude, duration, rates and coordination among locations can differ
even when selected dates agree. Retaining the fitted trajectory
alongside its dates preserves these properties for ecological analysis
and makes it possible to revisit transition definitions as the
ecological question changes. Where magnitudes are relevant, comparisons
can retain the original units within each response variable.

For ecological interpretation, canopy structure, physiological activity
and resource provision can be related through matched trajectories and
independent observations. Different vegetation indices may preserve
different aspects of development, producing different asymmetry
parameters and growth curve separations. Comparing those properties
offers a direct way to investigate which indicators best represent
foliage development, photosynthetic activity or other ecological
functions, and how their relationships change through the season.

\subsection{Further development and empirical
evaluation}\label{further-development-and-empirical-evaluation}

The analytical results define a tractable reference for extending the
work to observed vegetation trajectories. The overlap criterion is
derived for a symmetric composite curve and the third-derivative
approximation. Unequal growth curve rates, asymmetry, changing baselines
and phase-wise fitting provide natural next cases. Exact curvature can
be evaluated in the numerical units used by each implementation, while
the synthetic fitting study can be extended to irregular sampling, data
gaps and correlated errors. Profile likelihood or resampling would help
characterize constrained fits and the uncertainty associated with
merging landmarks.

An empirical evaluation can compare complete VI trajectories with
matched near-surface or canopy-development observations across short
seasons, asymmetric growth and disturbance. The analysis can test
whether different indices exhibit systematic differences in shape,
realized half-maximum width and growth curve separation, then assess the
corresponding landmark availability and date uncertainty. Klosterman et
al. (2014) provide an empirical precedent for combining curve shape,
observational scale and uncertainty; the present analytical
relationships offer testable explanations for the differences. Applying
both phase-wise and composite retrievals to the same observations would
connect those explanations to operational practice.

\section{Conclusions}\label{conclusions}

Logistic vegetation trajectories provide an explicit connection between
growth assumptions and phenological summaries. Building on the
ordinary-logistic threshold correspondence of Shang et al. (2017), the
analysis quantifies how asymmetry changes landmark thresholds, how
overlap changes seasonal amplitude and width, and when interior
third-derivative landmarks merge in a symmetric composite curve. Exact
curvature additionally depends on the selected axis metric. Local
sensitivity and nonlinear fitting link these geometric properties to
date uncertainty, showing that reduced bias from shape flexibility can
accompany increased uncertainty in the seasonal tails. These results
support retaining vegetation trajectories and reporting each date
together with its defining curve feature, amplitude reference and
estimation uncertainty.

\section*{Data and code availability}\label{data-and-code-availability}

No observational data were used. The accompanying reproducibility
archive contains the Python analysis script, software versions,
individual Monte Carlo results, scenario summaries and figure outputs.
The script uses NumPy, SciPy and Matplotlib. A permanent public
repository and identifier should be established before journal
submission.

An interactive web app, the Phenology Trajectory Explorer, is available
on GitHub Pages at
\url{https://bret-biodiversa.github.io/phenology-explorer.html}.

\section*{Notes on AI usage}\label{notes-on-ai-usage}

This manuscript grew out of two years of my thinking and preliminary
work on plant growth trajectories. My starting idea was that I found to
combine exponential cell division with a Michaelis--Menten relationship
(Michaelis and Menten, 1913) a generalized logistic representation of
canopy growth can be derived under explicit simplifying assumptions.

AI contributed substantially to developing this draft, including
mathematical elaborations, synthetic simulations, analytic examples,
code development and manuscript preparation. These contributions
extended my initial work and helped explore its implications. The
simulations are synthetic demonstrations, rather than analyses of my
satellite observations. I remain responsible for checking the
mathematics, code, interpretations and references.

There is still much to investigate. For example, which vegetation
indices most closely follow a generalized logistic trajectory, and how
does the shape parameter \(\alpha\) vary among indices and vegetation types?
Which indices show greater separation between the growth and senescence
growth curves? I initially called this separation ``FWHM season
length,'' although growth curve separation and the realized full width
at half maximum need to be distinguished.

I am enthusiastic about sharing this work on arXiv.org to support
further investigation and discussion. The contribution builds on
established logistic growth models and the landmark relationships
demonstrated by Shang et al. (2017), and develops their implications for
asymmetry, seasonal overlap and uncertainty. Empirical applications and
further theoretical development can now test and extend these
relationships.

\section*{References}\label{references}

\noindent\hangindent=1.5em\hangafter=1
Farquhar, G. D., von Caemmerer, S., and Berry, J. A. (1980). A
biochemical model of photosynthetic CO2 assimilation in leaves of C3
species. \emph{Planta}, 149, 78--90. \url{https://doi.org/10.1007/BF00386231}\par\medskip

\noindent\hangindent=1.5em\hangafter=1
Hikosaka, K. (2005). Leaf canopy as a dynamic system: ecophysiology and
optimality in leaf turnover. \emph{Annals of Botany}, 95, 521--533.
\url{https://doi.org/10.1093/aob/mci050}\par\medskip

\noindent\hangindent=1.5em\hangafter=1
Kawano, T., Wallbridge, N., and Plummer, C. (2020). Logistic Models for
Simulating the Growth of Plants by Defining the Maximum Plant Size as
the Limit of Information Flow. Plant Signaling \& Behavior, 15(2),
1709718. \url{https://doi.org/10.1080/15592324.2019.1709718}\par\medskip

\noindent\hangindent=1.5em\hangafter=1
Klosterman, S. T., Hufkens, K., Gray, J. M., Melaas, E., Sonnentag, O.,
Lavine, I., Mitchell, L., Norman, R., Friedl, M. A., and Richardson, A.
D. (2014). Evaluating remote sensing of deciduous forest phenology at
multiple spatial scales using PhenoCam imagery. Biogeosciences, 11,
4305--4320. \url{https://doi.org/10.5194/bg-11-4305-2014}\par\medskip

\noindent\hangindent=1.5em\hangafter=1
Michaelis, L., and Menten, M. L. (1913). Die Kinetik der
Invertinwirkung. \emph{Biochemische Zeitschrift}, 49, 333--369.\par\medskip

\noindent\hangindent=1.5em\hangafter=1
Moon, M., Zhang, X., Henebry, G. M., Liu, L., Gray, J. M., Melaas, E.
K., and Friedl, M. A. (2019). Long-term continuity in land surface
phenology measurements: A comparative assessment of the MODIS land cover
dynamics and VIIRS land surface phenology products. \emph{Remote Sensing
of Environment}, 226, 74--92. \url{https://doi.org/10.1016/j.rse.2019.03.034}\par\medskip

\noindent\hangindent=1.5em\hangafter=1
Richards, F. J. (1959). A flexible growth function for empirical use.
\emph{Journal of Experimental Botany}, 10, 290--301.
\url{https://doi.org/10.1093/jxb/10.2.290}\par\medskip

\noindent\hangindent=1.5em\hangafter=1
Shang, R., Liu, R., Xu, M., Liu, Y., Zuo, L., and Ge, Q. (2017). The
relationship between threshold-based and inflexion-based approaches for
extraction of land surface phenology. \emph{Remote Sensing of
Environment}, 199, 167--170. \url{https://doi.org/10.1016/j.rse.2017.07.020}\par\medskip

\noindent\hangindent=1.5em\hangafter=1
VIIRS GLSP Science Team (2016). \emph{Algorithm Technical Background
Document for VIIRS Global Land Surface Phenology}. NASA VIIRS Land.
\url{https://viirsland.gsfc.nasa.gov/PDF/VIIRS_GLSP-ATBD_V1.pdf}\par\medskip

\noindent\hangindent=1.5em\hangafter=1
VIIRS GLSP Science Team (2020). \emph{VIIRS Global Land Surface
Phenology Product User Guide}. Version 1.1, March 2020. Investigators X.
Zhang, M. A. Friedl and G. M. Henebry. Especially pp.~19--22.\par\medskip

\noindent\hangindent=1.5em\hangafter=1
Verma, M., Friedl, M. A., Finzi, A., and Phillips, N. (2016).
Multi-criteria evaluation of the suitability of growth functions for
modeling remotely sensed phenology. Ecological Modelling, 323, 123--132.
\url{https://doi.org/10.1016/j.ecolmodel.2015.12.005}\par\medskip

\noindent\hangindent=1.5em\hangafter=1
Zhang, X. (2015). Reconstruction of a complete global time series of
daily vegetation index trajectory from long-term AVHRR data.
\emph{Remote Sensing of Environment}, 156, 457--472.
\url{https://doi.org/10.1016/j.rse.2014.10.012}\par\medskip

\noindent\hangindent=1.5em\hangafter=1
Zhang, X., Friedl, M. A., Schaaf, C. B., Strahler, A. H., Hodges, J. C.
F., Gao, F., Reed, B. C., and Huete, A. (2003). Monitoring vegetation
phenology using MODIS. \emph{Remote Sensing of Environment}, 84,
471--475. \url{https://doi.org/10.1016/S0034-4257(02)00135-9}\par\medskip

\noindent\hangindent=1.5em\hangafter=1
Zhang, X., Liu, L., Liu, Y., Jayavelu, S., Wang, J., Moon, M., Henebry,
G. M., Friedl, M. A., and Schaaf, C. B. (2018). Generation and
evaluation of the VIIRS land surface phenology product. \emph{Remote
Sensing of Environment}, 216, 212--229.
\url{https://doi.org/10.1016/j.rse.2018.06.047}\par\medskip

\end{document}